\documentclass[aps,pra, twocolumn]{revtex4-1}
\usepackage{amsthm}

\usepackage[utf8]{inputenc}
\usepackage[sc,osf]{mathpazo}

\usepackage{verbatim} 
\usepackage{graphicx,subfigure}
\usepackage{bm}
\usepackage{makeidx} 
\usepackage{amsmath}
\usepackage{amssymb}
\usepackage{color}
\usepackage{amsthm}
\usepackage{hyperref}
\usepackage{cleveref}
\usepackage{diagbox}
\usepackage{physics}
\hypersetup{colorlinks, citecolor=magenta, filecolor=blue, linkcolor=blue, urlcolor=green}
\usepackage[normalem]{ulem}
\usepackage{float}
\usepackage{multirow}

\newcolumntype{P}[1]{>{\centering\arraybackslash}p{#1}}

\newcommand{\ph}[1]{{\color{red}#1}}

\begin{document}
\title{
%Limitation in recycling of states and disentangling of states\\
Sequential Reattempt of Telecloning}

\author{Sudipta Das,  Pritam Halder, Ratul Banerjee, Aditi Sen(De)}

\affiliation{ Harish-Chandra Research Institute, A CI of Homi Bhabha National
Institute,  Chhatnag Road, Jhunsi, Allahabad - 211019, India
	}

\begin{abstract}

The task of a telecloning protocol is  to send an arbitrary qubit possessed by a sender to multiple receivers. Instead of performing Bell measurement at the sender's node, if one applies unsharp measurement, we show that the shared state can be recycled for further telecloning protocol. 
Specifically, in case of a single sender and two receivers, the maximal attempting number, which is defined as the maximum number of rounds used by the channel  to obtain quantum advantage in the fidelity, turns out to be \emph{three} both for optimal and nonoptimal shared states for telecloning while the maximal number reduces to \emph{two} in case of three receivers.  Although the original telecloning  with quantum advantage being possible for arbitrary numbers of receivers, we report  that the recycling of resources is not possible in telecloning involving a single sender and more than three receivers, thereby demonstrating a no-go theorem.  We also connect the maximal achievable fidelities in each round with the  bipartite entanglement content  of the reduced state between the sender and one of the receivers as well as with the monogamy score of entanglement.

\end{abstract}

\maketitle

\section{Introduction}
\label{sec:intro}

Efficient information transmission  among distant parties  is one of the thriving avenues in the field of communication. Although  the existing communication protocols  serve our most of purposes, it has been realized  that the performance of these classical protocols can  be improved qualitatively by using quantum mechanical laws \cite{dc, tele@bennett, crypto1, AditiComm, pirandola_multiparty-tele_2015, Pirandola:20}. 
%Establishment of quantum communication network by means of genuine multipartite entangled state has been an active direction of research. 
Specifically, successful transmissions of classical and quantum information between a single  sender and a single receiver via quantum channels have been proposed and experimentally verified through various protocols like teleportation \cite{tele@bennett, tele-pan-photon}, dense coding \cite{dc, DC_obstacle, dcrev}, quantum key distribution \cite{BB84, crypto1, crypto2, Crypto3, Crypto4, Keydist}.  The possible next step is to generalize these protocols in a multiparty scenario, thereby building a communication network or quantum internet which is one of the centre of  attentions in recent years \cite{Kimble08}. In this direction,  prominent works include the measurement-based method for transmitting information over a long distance, known as quantum repeaters based on entanglement swapping \cite{zukowski93, Briegel98}, combination of cloning and teleportation to share an arbitrary quantum state between the senders and the receivers, called the  telecloning scheme \cite{ telecloning99, cloning98}, distributed quantum computing in a network \cite{distcomp99}, quantum dense coding network involving multiple senders and a single or two receivers \cite{bruss04}. Notice that in all these situations, shared multipartite entangled states are shown to be the key ingredient for successful realizations \cite{hhhh}. 

%connection between distant parties which are shown to be realized through intermediate nodes by the . In distributed quantum computing \cite{distcomp99} nonlocal quantum computational task was performed using multiparticle entangled states.

Despite giving  advantages during various information processing tasks over their classical counterparts, quantum mechanical laws also enforce stringent conditions on some available resources, recognized as no-go theorems \cite{nocloning, nobroadcast, nobitcomm, nodeleting, nomasking}. Among them, no-cloning theorem restrain us from copying an arbitrary quantum state perfectly 
%with the aid of universal quantum cloning machine 
\cite{nocloning, DIEKS1982271, YUEN1986405} although an approximate universal cloning machine exists  by which a quantum state can be copied with a certain fidelity \cite{buzek97, buzek98, cloning98}.
Taking optimal cloned state and performing teleportaion, information of a quantum state can be transferred from a sender to multiple receivers with optimal fidelity \cite{cloning98, telecloning99, Ghiu03, rigolin07} -- the protocol is known as telecloning which will be the main focus of this work. 
%Optimal cloning for $1 \rightarrow 2$, $N \rightarrow M$ realized in spin networks have been discussed in \cite{chiara04, chiara05, jiang07}. 
%In telecloning scenario, taking optimal cloning state and performing teleportaion, information of a quantum state can be transferred to multiple receivers with optimal fidelity \cite{telecloning98, telecloning99}. 
Instead of generating optimal clones \cite{buzek97,buzek98} and sending them to arbitrary number parties which require several bipartite entangled resource states, one can achieve the same task in  quantum telecloning protocol with less resources by a single measurement provided the optimal multipartite state is apriori shared between the senders and the receivers.  Since the protocol involves a projective measurement at the sender's side which destroys quantum correlations between the sender and the receivers, the shared state cannot be used for any other purpose in later time.  
%different story emerges when the single or the multiple receivers cannot perform local rotation, thereby unabling to complete the task. 

%$M$ number of initial $e$-bits and $M$ independent $2$-bit classical communication along with many extra qubits and two-qubit operations on Alice's side. On the contrary, perquisite of telecloning scheme is only $log_2 (M+1)$ $e$-bits through which Alice can transfer all $M$ clones with maximal possible fidelity by means of a single measurement on her side and two bits of classical messages \cite{telecloning99}.

%With the aim of maximizing the average fidelity of the telecloned states, the shared state can be used only once by projective measurement on Alice's node which destroys the quantum correlations between the sender and the receivers. 

\begin{figure}
    \centering
    \includegraphics[width=0.9\linewidth]{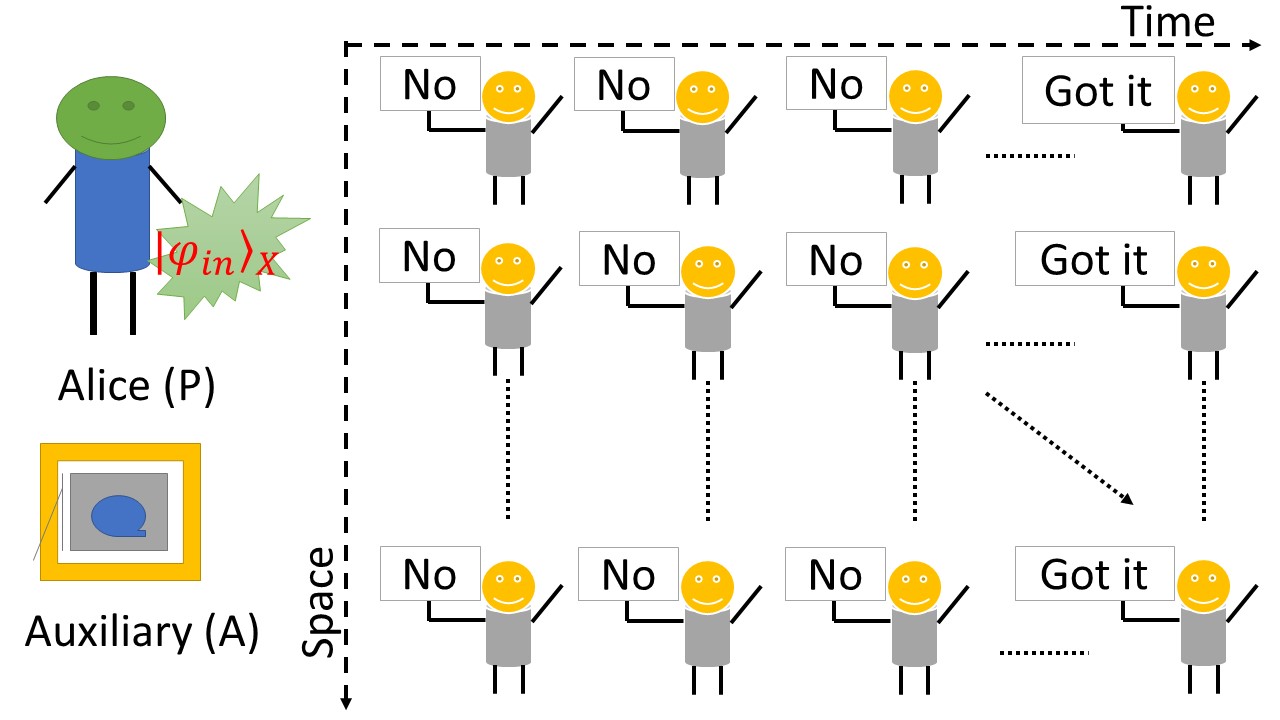}
    \caption{\textbf{Schematics of sequential telecloning protocol.} A single sender, Alice, called port, share an entangled state with $M$ Charus who do not perform the prescribed  rotation at their end for $n$ times. To send an arbitrary qubit, Alice performs a weak measurement which creates a possibility to reuse the state after Charus refusal for the next round of telecloning. We assume that \(M\) Charus are spatially separated and $n$ attempts to finish the protocol occurs at different time.   }
    \label{fig:protocol}
\end{figure}

%But there can be a different story where the single or multiple receivers doesn't complete the task of local rotation and state transfer is not complete. Then 
At this point, the natural question arises -- \emph{ for some reasons,  if the single or the multiple receivers do not complete the protocol,  can one design a protocol in such a way that the shared entangled state can be reused for  telecloning again}? In case of projective measurement performed by the sender, the answer is immediately negative. If we now assume that the measurement process is not perfect, i.e., instead of projective measurement, unsharp (weak) measurements are performed at the sender's side, the answer can be affirmative. We will now concentrate on the scenario where the shared state can be reused for the purpose of  telecloning (see Fig. \ref{fig:protocol} for schematics).

Using weak measurements, such sequential implementations of  protocols involving two parties have recently been observed in several directions  which include  violation of Bell inequalities \cite{Silva15, Mal16, Brown20, hu18,Schiavon_2017,feng20}, detecting entangled states with the help of steering inequalities \cite{sasmal18}, and entanglement witnesses \cite{Bera18},  the scenario of bi-nonlocal inequalities \cite{tavakoli21, tavakolinew, binonlocal-pritam22}, and reusing states for quantum teleportation \cite{ROY2021127143} to name a few.  Specifically, it was found that in these sequential scenarios,  %was first questioned in \cite{gisin15} and 
 at most two observers can share the nonlocality \cite{Silva15, Mal16} and  maximum twelve observers can witness entanglement at one side  \cite{Bera18} while maximum six receivers can reuse the shared state for teleportation \cite{ROY2021127143}. In  multipartite settings, a limited number of works has also been reported  in the direction of detecting multipartite entangled states \cite{Maity20, Gupta21, chirag22}.

%Accomplishment of this purpose is only possible by doing weak (unsharp) positive operator valued measurements (POVM) on sender's part such that in each round minimum disturbance of the resource ensures the fiedlity of the receiver's state to be greater than classical teleportation fidelity $2/3$. Maximum recycling number $(N_{Rec})$ of the observers during sequential exhibition of quantum correlations has been studied extensively in various circumstances. Sequential usage of the shared entangled state in the teleportation scheme was discussed in \cite{ROY2021127143}. % While the sequential scenario in bipartite entanglement scenario \cite{Bera18}, Bell nonlocality \cite{Mal16,gisin15}, binonlocality \cite{binonlocal-pritam22} are studied extensively we ask the same question in telecloning.
%Nonlocality of a single particle from an entangled pair can be shared for an arbitrary long sequence of independent observers \cite{gisin15} in the form of CHSH Bell inequality violation. 

We report here that by using optimal telecloned state,  
the shared state can be reused \emph{thrice}
%protocol can be implemented thrice
with the aid of weak measurement at the sender's port, when  two receivers are involved while 
the maximum number for attempting the scheme reduces to \emph{two}, when there are three receivers.
 Interestingly, we observe that for a fixed number of receivers, maximal attempting number remains unaltered even with the nonoptimal shared state,  provided the parameters are adjusted appropriately. Moreover, we find that recycling of state becomes impossible when there are more than three receivers. We also establish a relation between  the fidelity obtained in each round of the protocol and the entanglement content of the bipartite reduced state \cite{vidal2002} as well as with the monogamy score of entanglement \cite{coffman2000, ckw2, monorev}.
%We also analyze the attempting number  when the nonoptimal telecloned state is shared between the sender and the receivers. 

%Experimental demonstration of Bell-nonlocality was discussed in \cite{hu18,Schiavon_2017,feng20}. 

Sec. \ref{sec:picture} illustrates the scenario of recycling the shared multipartite state used in telecloning.
In Sec. \ref{sec:result3}, we present the main results, show the maximum number of recycling possible when the optimal  state for telecloning is shared between a single sender and two receivers  and establish a connection between entanglement content of the shared state and the fidelity obtained in this process while Sec. \ref{sec:multiple} deals with the recycling protocol involving a single sender and an arbitrary number of receivers. 
%how to recycle numbers for shared optimal state depending on each state's required fidelity. 
In contrast, Sec. \ref{sec:nonoptimal} shows the maximum recycling number even when the nonoptimal state  is shared between the sender and receivers. We summarize in Sec. \ref{sec:discussion}. 

%how disentangling affects the recycle number. In Sec. \ref{sec:rvb} we will demonstrate that telecloning is a state-dependent process.

%\textcolor{red}{We need a schematic + we need some figures to illustrate the protocol and optimized state vs non optimal state results. }

\section{Picture of recycling in network}
\label{sec:picture}

Teleportation is the transfer of an arbitrary qubit from a sender, Alice, to a receiver, Charu, with the help of two-bits of classical communication and a shared entangled state between Alice and Charu \cite{tele@bennett}.
When the shared state is maximally entangled, it is always possible to teleport an arbitrary qubit to Charu  while in case of a non-maximally entangled state,  the fidelity between an arbitrary qubit to be teleported and the state created at Bob's end has to be maximized after optimizing over  measurement at Alice's side and the rotation at Bob's part \cite{horodecki99}. 
%In this case, Alice and Bob have shared a maximally entangled state, $\ket{\zeta}=\frac{1}{\sqrt{2}}\left(\ket{00}+\ket{11}\right)$, and Alice wants to send her unknown qubit  $\ket{\phi_{in}}=\alpha\ket{0}+\beta\ket{1}$ to Bob.  First, Alice measure her qubits on the Bell basis and convey two bits of classical communication to Charu and according to that, Charu applies appropriate unitary operator. Thus using $1$ e-bit and $2$-bit classical channel any quantum state can be perfectly teleported to a distant party. 

%But, even though teleportation protocol is very faithful to sending unknown qubit, it has a limitation. Teleportation protocol is restricted between one sender-one receiver case.\\

Let us now consider a scenario involving a single sender and multiple receivers  where the sender, Alice, wants to send an unknown qubit to $M$ number of spatially separated receivers, Charus. We call the Alice's end as port, denoted by \(P\).  
Alice can  make multiple clones at her node locally and sends each of them to each  Charu via  teleportation protocol. However, this is not an efficient protocol and requires much more resource than the scenario when the shared state is multipartite entangled which is known as telecloning protocol. Suppose Alice and $M$ Charus, denoted by \(C_1, \ldots, C_M\)  share a multipartite entangled state, $\rho_{PC_{1}C_{2}...C_{M}}$
%For that Alice needs to use $M$ units of initial e-bits entangled channel and $M$ independent $2$ bit classical message which is  of course not efficient from a resource allocating view. This problem can be solved via exploiting a multipartite entangled state shared between sender and receivers where Alice only needs to publicly broadcast her measurement outcome. 
%This generalization of a teleportation protocol is called  telecloning.  
%Alice can simultaneously teleport her state to different receivers conclusively with an optimal fidelity, which is off course less than $1$ due to restriction from ``No-cloning'' \cite{} theorem.  
%Initially Alice and multiple Charus share a multiparty entangled state $\rho_{PC_{1}C_{2}...C_{M}}$ 
used as  a multi-receiver teleportation channel and the input state which is in possession with Alice to be teleported denoted as $|\phi_{in}\rangle$ at site \(X\). Alice performs a measurement in the Bell-basis, \(\{ |B_{1,2}\rangle = \frac{1}{\sqrt{2}}(|00\rangle \pm |11\rangle), |B_{3,4}\rangle = \frac{1}{\sqrt{2}}(|01\rangle \pm |10\rangle)\}\) jointly on her part of the entangled state and the input state.   Depending on the classical communication about  Alice's measurement outcome, different Charus perform their corresponding unitary operations at their node. In this way, Alice can simultaneously teleport the unknown qubit to different receivers conclusively with an optimal fidelity, which cannot reach to unity due to the  no-cloning  theorem \cite{nocloning} and is bounded above by the optimal fidelity obtained from approximate universal cloning machine \cite{buzek97, buzek98, cloning98}. 

We now present the entire protocol mathematically. Let us define a string $\{\small c\} = \{\small c_{1},\small c_{2}, \ldots \small, c_{M}\} $  where $\small c_{i}= 1 $ if \(i\)th Charu $C_{i}$ is participating in the protocol, and applies local unitary while $\small c_{i}= 0 $, otherwise. If the initial state can be represented as
\begin{eqnarray}
    & \rho_{in} &= |\phi_{in}\rangle_{X}\langle\phi_{in}|\otimes\rho_{PC_{1}C_{2}\ldots C_{M}}, 
    \end{eqnarray}
   the reduced state at one of the Charu's end,  $i$, after the  measurement and applying the corresponding  unitary operation, can be written as   
    \begin{eqnarray}
    \rho_{{C_{i}}} &= T^{i}(\rho_{in}) = Tr_{\{\overline C_{i},X,P\}}(\sum _{k} \mu_{k} \rho _{in} \mu^{\dagger}_{k})\ph{,} \\\nonumber %&&F^{i}(\rho_{AC_{1}..C_{M}}) &=& \int \langle \phi_{in}|T_{i}(\rho_{AC_{1}..C_{M}})|\phi_{in}\rangle d\phi_{in}\\ \nonumber
\end{eqnarray}
where $\overline C_{i} = \bigcup_{j=1, j \neq i}^{M}C_{j} $ and $ \mu_{k}= \sqrt{\mathcal{M}_{k}}\otimes_{j=1}^{M}(U_{k})^{c_{j}}$ with $\sum_{k}\mathcal{M}_{k} = \mathbb{I}$  being the set of positive operator valued measurements at Alice's part. In the projective Bell-measurement scenario, $\mathcal{M}_{k} =\ket{B_k}\bra{B_k}$ and the set of unitaries  at node $i$ for the output is 
%\begin{eqnarray}
 % &\ket{B_i}& = 
    %    \begin{cases}
     %       \frac{1}{\sqrt{2}}\left(\ket{00}-\left(-1\right)^i\ket{11}\right) & \text{if $i=1,2$}\\\nonumber
      %      \frac{1}{\sqrt{2}}\left(\ket{01}-\left(-1\right)^i\ket{10}\right) & \text{if $i=3,4$}.
    %    \end{cases} \\\nonumber
\(\{\mathbb{I},\sigma_{x},\sigma_{y},\sigma_{z}\}\).  %\nonumber 
%\label{fidelity1}
%\end{eqnarray}
For a particular channel used between Alice and multiple Charus, the average fidelity of a teleported state at  $i$-th receiver's node over all possible input states can be defined as
\begin{eqnarray}
  &F^{i}(\rho_{PAC_1 \ldots C_M}) &=\int \langle \phi_{in}|T^{i}(\rho_{in})|\phi_{in}\rangle d\phi_{in}.\\ \nonumber
\end{eqnarray}
Notice that the fidelity obtained here is same even when the measurement performed at Alice's port is the unsharp measurement. 

For the  telecloning protocol, we require to replace the initial multiparty state between Alice and Charus, with a particular type of multiparty entangled channel  from the universal optimal cloning machine \cite{telecloning99, cloning98}. 
We choose the shared $2M$-partite entangled resource state to be 
\begin{eqnarray}
\ket{\psi}_{PAC}=\frac{1}{\sqrt{2}}\left(\ket{0}_P\ket{\phi_0}_{AC}+ \ket{1}_P\ket{\phi_1}_{AC}\right),
\label{state1}
\end{eqnarray}
where
\begin{eqnarray}
\nonumber \ket{\phi_0}_{AC}=\sum_{j=0}^{M-1}\alpha_j\ket{A_j}_A\otimes\ket{\{0,M-j\},\{1,j\}}_C, \\\nonumber
\ket{\phi_1}_{AC}=\sum_{j=0}^{M-1}\alpha_j\ket{A_{M-j-1}}_A\otimes\ket{\{0,j\},\{1,M-j\}}_C,\\ 
\label{state2}
\end{eqnarray}
with
\begin{eqnarray}
    \nonumber \ket{A_{j}} = \ket{\{0,M-j-1\},\{1,j\}} \\
    \alpha_j=\sqrt{\frac{2\left(M-j\right)}{M\left(M+1\right)}}.
\end{eqnarray}
%Here, $P$ signifies the "port" qubit held by Alice and 
Here $\ket{\{0,a\},\{1,b\}}$ is the symmetric and normalized state of $(a+b)$ qubits with '$a$' no. of qubits in state $\ket{0}$ and remaining '$b$' no of qubits are in orthogonal state $\ket{1}$. 
$A$ refers to  the $M-1$ qubit auxiliary system, which is also at Alice's side by convention, although it can be at a different location. In this case, the shared state between Alice and Charus reads as
%Also, $C$ includes all the $M$ receivers $C_1$, $C_2,...,$ $C_M$. Therefore, tensor product of the unknown state $\ket{\phi_{in}}_X$ with the shared resource state can be written as
\begin{eqnarray}
\label{initial}
\rho_{in} = \ket{\phi_{in}}_{X}\bra{\phi_{in}}\otimes\rho_{PAC_{1}\ldots C_{M}}\nonumber\\
\rho_{PAC_{1}..C_{M}}= \ket{\psi}_{PAC_{1}..C_{M}}\bra{\psi}. 
\end{eqnarray}
Let us now discuss the sequential scenario of the telecloning protocol. 
\begin{itemize}
    \item Suppose  only $M^{\prime}$ number  of Charus agree to apply their unitary operations and receive the teleported states.  In this situation, we define a map to get back the  recycled channel of $(2M - M^{\prime}) $-party state, where  $M'$ charus  are traced out and finally averaging is performed over uniformly generated  input states so that the recycled channel does not depend on a particular input state. Note that  finally averaging over the input states is equivalent to initially taking the input state as the average state $\int |\phi_{in}\rangle\langle\phi_{in}|d\phi_{in} = \frac{\mathbb{I}}{2}$ \cite{ROY2021127143}.
   The recycled channel, in this case, is given by
\begin{eqnarray}
\nonumber &R^{\{c\}}&({\rho_{PAC_{1} \ldots C_{M}}}) \\\nonumber &=& \int Tr_{\{\{ C_{i}\forall i ; c_{i}=1\},X\}}(\sum _{i} \mu_{i} \rho _{in} \mu^{\dagger}_{i})     d\phi_{in}\\\nonumber
&= & Tr_{\{\{ C_{i}\forall i ; c_{i}=1\},X\}}(\sum _{i} \mu_{i} ((\int |\phi_{in}\rangle\langle\phi_{in}|d\phi_{in}) \otimes  \rho_{PAC_{1} \ldots C_{M}})  \mu^{\dagger}_{i} \\
&=&  Tr_{\{\{ C_{i}\forall i ; c_{i}=1\},X\}}(\sum _{i} \mu_{i} (\frac{\mathbb{I}}{2} \otimes  \rho_{PAC_{1} \ldots C_{M}}) \mu^{\dagger}_{i}. 
\label{recyc1}
\end{eqnarray}
Note that the map defined above does not act on the auxiliary system, i.e., identity operators only act on them.

    \item Depending on  the type of measurement performed on Alice's part, the recycled channel becomes useful for teleportation in the next round.
    %For projective Bell measurement , the recycled channel will not have any correlation left. 
    Instead of projective measurement, if Alice performs  an unsharp Bell measurement,  given by
    %as shown below in the Alice's part allow us to recycle the channel in such a way that can be useful in further rounds. 'Telecloning' protocol allowed us the generalization of classic teleportation in  space-like separated multi-receiver scenario. In our work, we show that using unsharp measurement formalism and introducing the recycling of quantum channel, sender will also be able to teleport her state to different time-like separated receivers, thus forming quantum communication network between Alice and different time-like separated Charus[fig.1]. All the previously introduced maps change accordingly. 
    \begin{eqnarray}
    &\mathcal{M}^{\lambda}_i&=\lambda \ket{B_i}\bra{B_i}+\frac{1-\lambda}{4}\mathbb{I}_4,
    %\\\nonumber
   % &\mu^{\lambda}_{i}&= \sqrt{\mathcal{M}^{\lambda}_{i}}\otimes_{i=1}^{M}(U_{i})^{c_{i}}
    \label{wmeasure}
    \end{eqnarray}
where \(\lambda\) is the unsharp parameter, we will show that the channel can further be used for more rounds depending on the residual entanglement. 
 Similarly, we can redefine, $R^{\{c\},\lambda}, F^{i,\lambda}, T^{i,\lambda}$ by replacing $\mu_{i}$ with $\mu^{\lambda}_{i} = \sqrt{\mathcal{M}^{\lambda}_{i}}\otimes_{j=1}^{M}(U_{i})^{c_{j}}$ in previously defined $R^{\{c\}}, F^{i},$ and $T^{i}$ consecutively.
  
  \item  Suppose upto round $n-1$, all the receivers refuse to collaborate in the telecloning protocol. The recycled channel through $n - 1$ round can be reused in the next round, $n$ and the corresponding average fidelity in the round $n$ can be calculated as
  \begin{eqnarray}
   \nonumber &F^{i,\lambda_{n}}&(\rho'_{PAC_1 \ldots C_M})= \\ \nonumber 
   &F^{i,\lambda_{n}}&(R^{\{0\},\lambda_{n-1}}\cdot R^{\{0\},\lambda_{n-2}}\cdot\cdot\cdot R^{\{0\},\lambda_{1}}\cdot(\rho_{PAC_1\ldots C_M})),\\
   \label{fidelity2}
    \end{eqnarray}
   with the definition $\{0\}= \{0,0,\ldots, 0\}$ where all  $c_i = 0$ and $\lambda_{i}$ is the unsharp parameter of Alice's measurement in the round $i$.

   \item Let us assume that in previous $(n-1)$ rounds, not all Charus are refusing. Hence the bit string of the round, $k$, $\{c\}_{k}$  with the information which Charus have refused , the recycling map acts accordingly and we get the required fidelity in the round $n$ as
    \begin{eqnarray}
  \nonumber &F^{i,\lambda_{n}}&(\rho'_{PAC_1 \ldots C_K})= \\ \nonumber
   &F^{i,\lambda_{n}}&(R^{\{c\}_{n-1},\lambda_{n-1}}\cdot R^{\{c\}_{n-2},\lambda_{n-2}}\cdot\cdot\cdot R^{\{c\}_{1},\lambda_{1}}\cdot(\rho_{PAC_1\ldots C_M}))\\
   \label{fidelity3}
   \end{eqnarray}
   where $K<M$ and total $(M-K)$ receivers receive the state in previous $(n-1)$ rounds and go out of the protocol in further round.
  We will evaluate all these situations in the succeeding section for a given state. 
 % &&K =  M - \mathcal{C}(S) ;  S = \{i\forall c_{i}=1\}
\end{itemize}

\section{Reattempting via optimal telecloned state}
\label{sec:result3}

In this section, we will mainly concentrate on the sequential telecloning protocol which starts with a tripartite entangled state shared between a single sender and two receivers along with the auxiliary state. After the unsharp measurement by Alice, we consider two situations -- (1) when both the Charus do not perform the unitary operations for a few rounds, (2) when one of the Charus wishes to finish the protocol while the other one refuses. We are also able to connect the fidelity obtained in each round with the entanglement content of the state in that round.

\subsection{Sequential telecloning with a single sender and two receivers }

Let us illustrate this protocol for the simplest scenario having a single sender and two receivers.
From Eq. (\ref{state1}), the optimal  state in this case
%between one sender and two receivers 
can be written as
\begin{eqnarray}
\label{optimalstate}
    \ket{\psi}_{PAC_1C_2}=\frac{1}{\sqrt{2}}\left(\ket{0}_P\ket{\phi_0}_{AC_1C_2}+ \ket{1}_P\ket{\phi_1}_{AC_1C_2}\right), \nonumber \\
\end{eqnarray}
where
\begin{eqnarray}
 \ket{\phi_0}_{AC_1C_2} &= &\sqrt{\frac{2}{3}}\ket{000}_{AC_1C_2}+\sqrt{\frac{1}{6}}\ket{101}_{AC_1C_2} \nonumber \\
    & + &\sqrt{\frac{1}{6}}\ket{110}_{AC_1C_2},
\end{eqnarray}
and 
\begin{eqnarray}
\ket{\phi_1}_{AC_1C_2} &= & \sqrt{\frac{2}{3}}\ket{111}_{AC_1C_2}+\sqrt{\frac{1}{6}}\ket{001}_{AC_1C_2}\nonumber \\
&+&\sqrt{\frac{1}{6}}\ket{010}_{AC_1C_2}.
\end{eqnarray}
 The  state
 %and  the quantum channel 
 between  the sender, \(P\), Charus, \(C_1\) and \(C_2\)  and the auxillary system along with the state to be teleported can be represented as
\begin{eqnarray}
\rho_{in} &=& |\phi_{in}\rangle_{X}\langle\phi_{in}|\otimes\rho_{PAC_{1}C_{2}}, \\\nonumber
 \mbox{where}\, \, \rho_{PAC_{1}C_{2}} &=& |\psi\rangle\langle\psi|_{PAC_1 C_2},\\\nonumber
\mbox{and} \,\,  |\phi_{in}\rangle_{X}&=&\alpha\ket{0}+ \beta\ket{1}. \\\nonumber
\end{eqnarray}
After the first round of unsharp measurement in Eq. (\ref{wmeasure}), the teleported state at  any one of the receiver's side, say \(C_1\), reduces to
\begin{eqnarray}
  \rho_{{C_{1}}} &=& T^{1,\lambda}(\rho_{PAC_{1}C_{2}}) = Tr_{\{ C_{2},X,P,A\}}(\sum _{i} \mu_{i}^{\lambda} \rho _{in} \mu^{\lambda \dagger}_{i}) \\\nonumber  &=& \begin{pmatrix}
\frac{1}{2}+\frac{\lambda}{3}\left(|\alpha|^2-|\beta|^2\right) & \frac{2}{3}\alpha\beta^*\lambda \\
\frac{2}{3}\alpha^*\beta\lambda & \frac{1}{2}+\frac{\lambda}{3}\left(|\beta|^2-|\alpha|^2\right) 
\end{pmatrix}\\
 &=& \frac{2}{3}\lambda\ket{\phi_{in}}\bra{\phi_{in}}+\frac{3-2\lambda}{6}\mathbb{I}_2.
\end{eqnarray}
The shared state is symmetric in both the receiver's end, and hence the second Charu, \(C_2\) also obtains the same state. Therefore,  after the first round, the expression of required fidelity as a function of the sharpness parameter $\lambda$ can be computed for a receiver, say, \(C_1\) as
\begin{eqnarray}
\label{eq:1s2r}
f_{1}&=& F^{1,\lambda}(\rho_{C_{1}})\\\nonumber
&=&\int\bra{\phi_{in}} T^{1,\lambda}(\rho_{in}) \ket{\phi_{in}}d\phi\\\nonumber
&=& \int\bra{\phi_{in}}\rho_{C_{1}}\ket{\phi_{in}}d\phi\\\nonumber
&=&\frac{1}{2}+\frac{\lambda}{3}.
\end{eqnarray}
It clearly demonstrates that there exists a range of \(\lambda\) above which any arbitrary state can be telecloned to both the receivers with a  fidelity more than the classical one, i.e., \(2/3\) \cite{massarpopescu}  while the maximal fidelity is in accordance with optimal cloning machine \cite{buzek97, buzek98, cloning98}, i.e.,  $f_1=\frac{5}{6}$ is achieved when the measurement is projective.  Notice that, instead of performing unsharp measurement in this scenario, the fidelity is a state-independent quantity since it is based on the  universal cloning. It implies that we do not require to consider average fidelity.
%, in accordance with the maximal fidelity achievable by the optimal cloning machine \cite{buzek97, buzek98, cloning98}.

\subsubsection{Unable to complete the protocol by both the  receivers}

%Now, we want to check whether by the
Let us now consider the situation when both \(C_1\) and \(C_2\) refuse to perform the corresponding unitary operations required to complete the protocol. Since unsharp measurement is performed at \(P\)'s node,  
%refusal of both Charus at receiver's end telecloning can be done in the next round by recycling 
even after the refusal, the shared resource state can possibly be used for another round of telecloning protocol. 
%Suppose alice performed her measurement but both of Charu did refuse to receive.
As discussed before, the optimal channel has to be recycled and to be used in the second round. The average fidelity for $C_{1}(C_{2})$ in this round can  be calculated as
\begin{eqnarray}
\label{eq:2s2r}
\nonumber
f_{2}&=&F^{1,\lambda_{2}}(R^{\{0\},\lambda_{1}}(\rho_{PAC_{1}C_{2}}))\\\nonumber&=& F^{1,\lambda_{2}}(\sum_{i}\mu^{\lambda_{1}}_{i}(\frac{\mathbb{I}}{2}\otimes\rho_{PAC_{1}C_{2}}) \mu^{\lambda_{1} \dagger}_{i})\\
&=& \frac{1}{2} + \frac{P(\lambda_{1})}{3}\lambda_{2} \\\nonumber
\mu^{\lambda_{1}}_{i} &=& \sqrt{\mathcal{M}^{\lambda_{1}}_{i}}\otimes_{i=1}^{M}(\mathbb{I}).
\end{eqnarray}
The subscript, \(i\) in \(\lambda_i\) denotes the round in which measurement is performed.
In a similar fashion, if both the Charus refuse to finish the process  till  round \(n-1\),  the average fidelity of $C_{1}(C_{2})$ in  the round \(n\) is found to be
\begin{eqnarray}
\nonumber
   &&f_{n}\\\nonumber&=& F^{1,\lambda_{n}}(R^{\{0\},\lambda_{n-1}}\cdot R^{\{0\},\lambda_{n-2}}\cdot\cdot\cdot R^{\{0\},\lambda_{1}}\cdot(\rho_{PAC_{1}C_{2}}))\\
   &=&\frac{1}{2}+\frac{P\left(\lambda_1\right)P\left(\lambda_2\right).....P\left(\lambda_{n-1}\right)}{3}\lambda_n,
   \label{both-refuse}
\end{eqnarray}
where $P\left(\lambda\right)=\frac{1}{2}\left[1-\lambda+\sqrt{\left(1-\lambda\right)\left(1+3\lambda\right)}\right]$. Note here that with the weakness parameter $\lambda_{i}$, the action of the recycled map, $R^{\{0\}, \lambda_{i}}$ depends on the predecided fidelity in the previous rounds. It is due to the fact that Alice performs the measurement having a prefixed \(\lambda\) value  according to this predecided fidelity, after which Charus refuse to act. By fixing the fidelity,  ${f_{i}} >2/3$,  we can calculate the range of unsharp parameter $\lambda_{i}$ for  $i= 1,\ldots ,n-1$. If one  demands the high fidelity in the previous round, the weakness parameters also approaches to unity, thereby reducing the quantum correlations in the recycled channel.
%close to 1 and Alice's measurement reduces the correlation in the recycled channel more.\\

\begin{figure}[ht]
\includegraphics[width=0.9\linewidth]{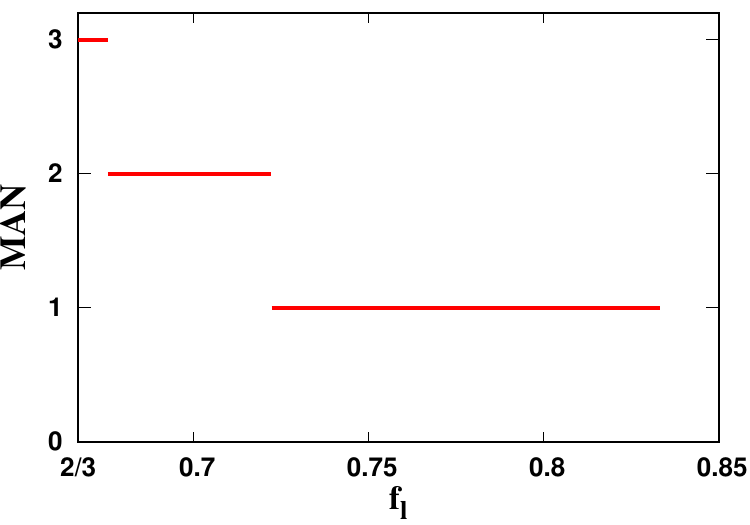}
\caption{(Color online.) Maximal attempting number (ordinate) vs. fidelity $(f_l)$. When we demand that each round has fidelity just above the classical one, the maximal attempting number by Charus becomes three while it decreases with the increase of the fidelity in each round.  Both the axes are dimensionless.  }
\label{fig:my_label} 
\end{figure}

\emph{Maximum attempting number.} Let us now define  the maximal number of rounds that  a channel can be used such that  the quantum advantage in the fidelity can be obtained  -- we call it as the maximum attempting number (MAN). It is well known that the classically achievable bound in teleportation is $f_{cl} = \frac{2}{3}$ \cite{massarpopescu} and hence  quantum enhancement is guaranteed when the fidelity is above \(\frac{2}{3}\).
If we demand the lower bound of fidelity in each round to be  $f_{i} \geq f_{l} \forall i $, we will reach to a round, $n_{cr}$ for which 
\begin{eqnarray}
&&0<\lambda_{i}\leq 1 \,\,  \forall i \in \{1,2,...,n_{cr}-1\},\\\nonumber
&& \mbox{and} \,\,  \lambda_{n_{cr}}>1. \nonumber
\end{eqnarray}
Therefore,  the round $(n_{cr}-1)$ signifies the maximal attempting number since 
%measurement corresponding to 
$\lambda>1$ is not a valid measurement. E.g., let us fix $f_{l}=0.675$. To satisfy this, we get a range of possible \(\lambda\) values in each round which leads to a fidelity more than \(f_l\) for a shared state in Eq. (\ref{optimalstate}), i.e.,  $\lambda_{1} \geq 0.525$, $\lambda_{2} \geq 0.664158$, $\lambda_{3} \geq 0.992511$ but $\lambda_{4} >1$. 
%=11.0228$.
 In Table.~\ref{tab:2refusal} we report the relation between this lower bound in fidelity  $f_{l}$  and the MAN for the optimal telecloned state (see Fig. \ref{fig:my_label}).
\begin{table}[H]
\begin{center}
\begin{tabular}{|c|c|}
\hline
     Range of $f_{l}$  & $MAN$ \\\hline
      $0.6667-0.6754$& 3\\\hline
      $0.6755- 0.7222$& 2\\\hline
      $0.7223-0.8333$&1\\\hline
\end{tabular}
\caption{Maximal attempting  number, MAN, when the fidelity of each round is greater than or equal to $f_{l}$.}
\label{tab:2refusal}
\end{center}
\end{table}

\subsubsection{Completion of protocol by a single receiver}

Let us now assume an asymmetric situation, i.e.,  any one of the receivers,  say $C_1$, performs the unitary operation communicated by the sender, thereby finishing the telecloning task while $C_2$ does not finish the protocol. We 
%are trying to answer the question,
will now address the question whether the acceptance by $C_{1}$ can have any affect on the reuseability of the channel with respect to Alice and $C_{2}$. We find that the answer is negative, i.e., reusability of the channel between port and $C_{2}$ does not depend on whether $C_{1}$ has completed its telecloning protocol or not.

In this picture, 
%ne of the receiver who refused in the first round applied no unitary in his part,but 
$C_{1}$ applies the unitary accordingly and leaves the protocol while \(C_2\) does not perform the unitary.
Hence in the second round, 
%So, in the next step , 2 receiver
the entire multipartite protocol reduces to a  standard teleportation with a single sender-receiver pair.
%receiver - one sender teleportation channel. 
The fidelity achieved by $C_{1}$ can be calculated to be same as before, i.e., 
\begin{eqnarray}
\nonumber
f^{C_{1}}_{1}&=&F^{1,\lambda_{1}}(\rho_{PAC_{1}C_{2}})\\
&=& \frac{1}{2} + \frac{\lambda_{1}}{3}. 
\label{eq:fid1}
\end{eqnarray}

Using the reduced scenario involving single sender - single receiver teleportation channel, $C_{2}$ can achieve fidelity in the  round \(n\) as
\begin{eqnarray}
\nonumber
&f^{C_{2}}_{n}&\\\nonumber
&=& F^{2,\lambda_{n}}(R^{\{0\},\lambda_{n-1}}\cdot R^{\{0\},\lambda_{n-2}} \cdot \cdot R^{\{c\}_{1},\lambda_{1}}(\rho_{PAC_{1}C_{2}}))\\
&=& \frac{1}{2}+\frac{P\left(\lambda_1\right)P\left(\lambda_2\right).....P\left(\lambda_{n-1}\right)}{3}\lambda_n,
\label{single-refuse}
\end{eqnarray}
where $\{c\}_{1}=\{1,0\}$ carries information that $C_{1}$ has  finished while
%the state in the first round but 
$C_{2}$ has not. Comparing Eqs. (\ref{both-refuse}) and (\ref{single-refuse}), we can confirm that the maximum attempting number
still remains three provided that each round has fidelity just above the classical one. 
Moreover, we notice that the achievable fidelity by $C_{2}$ in each round does not depend on $C_{1}$'s refusal or acceptance on the first round.

\subsection{Connecting entanglement with fidelity in a sequential scenario}

\begin{figure}[ht]
\includegraphics[width=0.9\linewidth]{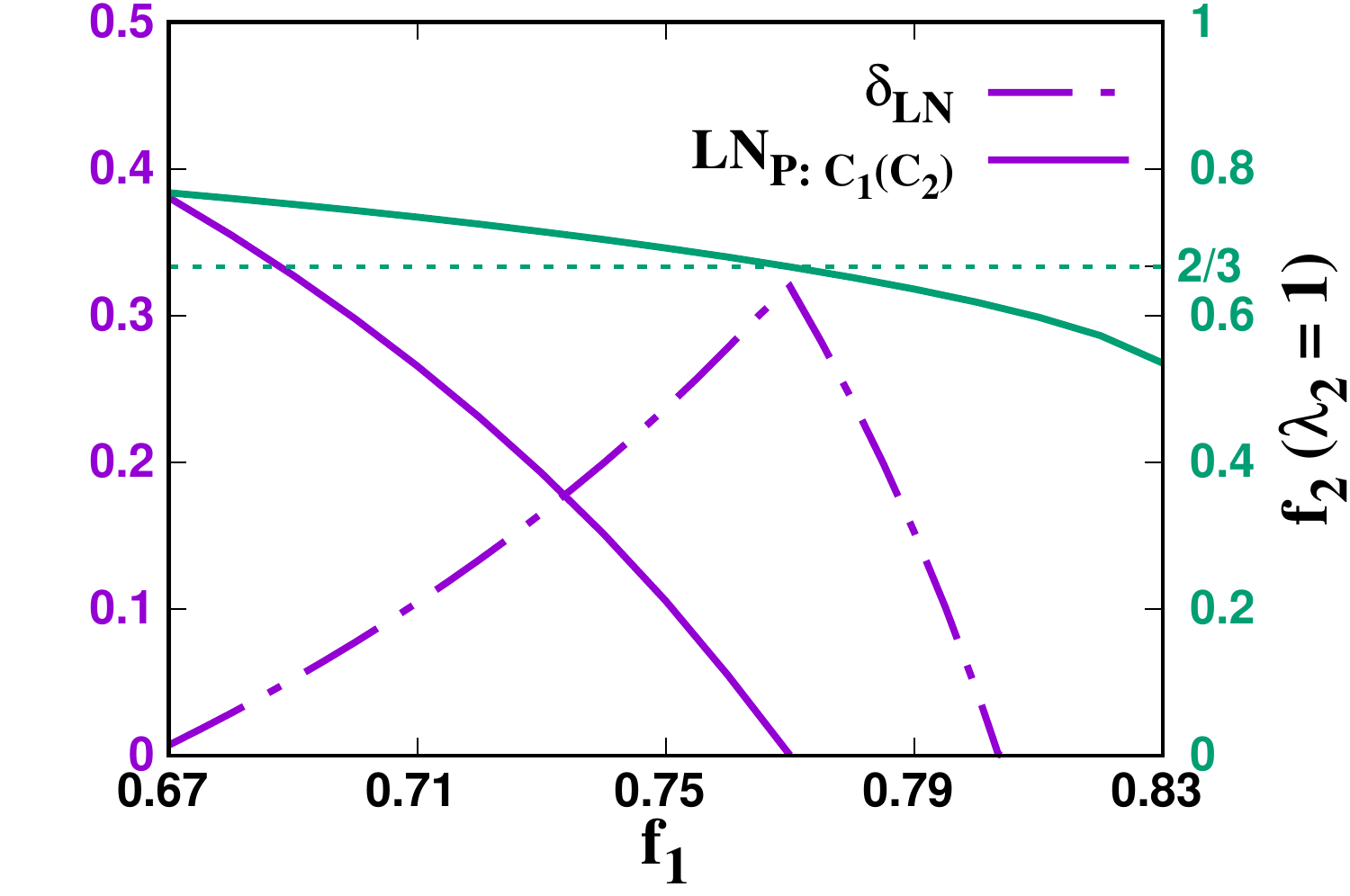}
\caption{(Color online.) Maximum achievable fidelity, \(f_2\)  (right ordinate) with $\lambda_{2}=1$, i.e., with the projective measurement in the second round against \(f_1\) (abscissa) in the first round of the protocol.  Entanglement in the bipartition \(LN_{P:C_1(C_2)}\)(left ordinate, solid line)  and monogamy score, \(\delta_{LN}\) (left ordinate, dashed line) defined in Eq. (\ref{eq:monogamysc}) obtained in the second round, i.e., of the first recycled state shared between port, auxiliary system and two receivers with respect to \(f_1\) (abscissa). Here we assume that in the first round, unsharp measurement is performed and both the Charus have not performed the unitary operations. 
All the axes are dimensionless. }
\label{fig:1st recycle} 
\end{figure}

\begin{figure*}[ht]
\includegraphics[width=\textwidth]{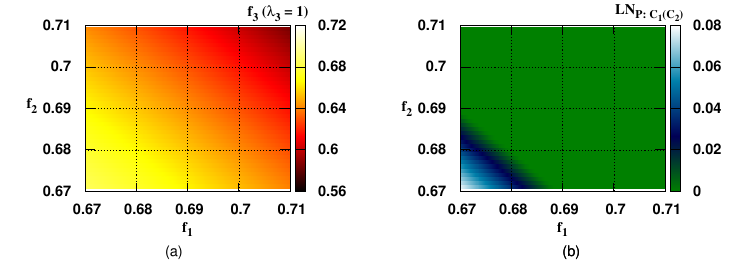}
\caption{(Color online.) Features in the second recycled state. (a) Map plot of the fidelity, $f_3$ with $\lambda_3=1$ in the plane of \(f_1-f_2\) (horizontal-vertical axis). (b) The behavior of $LN_{P:C_1(C_2)}$ of the second recycled shared state with \(f_1\) (abscissa) and \(f_2\) (ordinate) which can be used to perform the teleportation in the third round. Both the axes are dimensionless.}
\label{fig:2nd recycle} 
\end{figure*}

We will establish a connection between entanglement content of the shared state used in each round and the fidelity obtained in that round. To quantify bipartite entanglement, we choose logarithmic negativity (LN) in Eq. (\ref{eq:LN}) for the reduced state between the sender (port) and one of the receivers. On the other hand, we also relate the shareability of entanglement in each round which we characterize via  monogamy of entanglement (see Appendix \ref{sec:monogamy}).

Let us first calculate the entanglement after the unsharp measurement is performed by the sender in the first round and both \(C_1\) as well as \(C_2\) decline to perform the unitary operations. In this situation, the fidelity, \ph{\(f_2\)}  is given in Eq. (\ref{eq:2s2r}). For a fixed fidelity,  the corresponding entanglement of the first recycled state between the port and one of the receivers, say \(C_1\) can be computed, \(LN_{P:C_1}\) after tracing out the  auxiliary qubit, and \(C_2\). It takes the form as
\begin{eqnarray}
    \nonumber && LN_{P:C_1}=\\
  \nonumber  &&\log_2(\frac{1}{6}|(0.5+3f_1-\sqrt{2.5-3f_1}\sqrt{9f_1-3.5}-2\sqrt{2}\\
     &\times&\sqrt{(2.5-3f_1)(3f_1-0.5+\sqrt{2.5-3f_1}\sqrt{9f_1-3.5})})|\nonumber\\
    &+&1). 
    \label{eq:LN1st}
    \end{eqnarray}
We find that by demanding  high fidelity in the first round, the entanglement in the recycled state decreases with the increase of \(f_1\) and vanishes for a certain \(f_1\) value, i.e., \(f_1 = 0.7697\) (as shown in Fig. \ref{fig:1st recycle}). If one uses this recycled state to perform another telecloning scheme with the projective measurement at the port's end,  the fidelity also decreases and goes  below the classical limit at the same point where entanglement vanishes as expected. It also illustrates that the entanglement content of the recycled channel in the first round is not enough to achieve the required fidelity, say, $f_2=0.7233$ in the second round. It also shows that entanglement is necessary but not sufficient to the successful implementation of the sequential telecloning protocol. 

Let us now analyze the monogamy score of LN, \(\delta_{LN}\) after the first round by taking port, $P$ as the nodal observer. It specifies the distribution of entanglement between different sites with respect to the port. We observe that \(\delta_{LN}\) actually reaches maximum at the same point where \(LN_{P:C_1}\) vanishes as depicted in Fig. \ref{fig:1st recycle}.

We now examine the situation after the second round. In this case, in both  first and  second rounds, unsharp measurements are performed and Charus do not perform the unitary operations. If we now study the behavior of entanglement of the recycled state between the port and \(C_1\) or  \(C_2\) in Eq. (\ref{eq:LN2}), we find that it is nonvanishing only when we demand fidelities, \(f_1\) and \(f_2\) to be just above the classical bound in  previous rounds, thereby giving fidelity in the third round beyond \(2/3\) (comparing Figs. \ref{fig:2nd recycle} (a) and (b)).

\section{No-go theorem for Recycling of telecloning with multiple receivers }
\label{sec:multiple}

Let us now move to the telecloning situation which involves a single sender and arbitrary number of receivers, say $M$. When the projective measurement is allowed at port's end, the fidelity of the telecloned state gradually decreases with the  increase of the number of  receivers, $M$ and for $M\to\infty$, the optimal average fidelity goes to $2/3$, the classical limit. This result suggests that  the opportunity to recycle the shared entangled state should also decrease with the increase of $M$ in case of  unsharp measurement at the sender's side.  The question that we address here --  what is the maximum number of receivers allowed, i.e., the maximum  $M$ upto which the recycling can happen? 

When \(M\) receivers refuse to finish the protocol in all the rounds till \(n-1\), the average fidelity at the round \(n\)  with \(M\) number of receivers can be computed as
\begin{eqnarray}
\label{fidelity for M receriver}
f_n=\frac{1}{2}+\left[\frac{M+2}{6M}\right]P\left(\lambda_1\right)P\left(\lambda_2\right).....P\left(\lambda_{n-1}\right)\lambda_n.
\end{eqnarray}
%\textcolor{red}{It has to be checked for arbitrary receiver, checked for $3$receiver}
%where $T_M=T_{M-1}+2M+3$ and $T_1=2$. \textcolor{red}{Here, $2M+ 3$  represents the number being added in the previous term. bujhlum na? } 
%The coefficient can be obtained from the approximate cloning machine, \(1 \rightarrow M\). 

Let us now elaborate the picture when there are three receivers along with a sender. 
%\begin{itemize}
The teleported state at one of the receiver's end, say \(C_1\) after the first round with $M=3$ looks like
    \begin{eqnarray}
  \nonumber  \rho_{C_1}
    &=&\begin{pmatrix}
\frac{1}{2}-\frac{5}{18}\lambda+ \frac{5}{9}\lambda|\alpha|^2 & \frac{5\lambda}{9}\alpha\beta^* \\
\frac{5\lambda}{9}\alpha^*\beta & \frac{1}{2}-\frac{5}{18}\lambda+ \frac{5}{9}\lambda|\beta|^2
\end{pmatrix}\\
    &=& \frac{5}{9}\lambda\ket{\phi_{in}}\bra{\phi_{in}}+\left(\frac{1}{2}-\frac{5}{9}\lambda\right)\mathbb{I}_2,
    \end{eqnarray}
and $f_n$ at the round \(n\) can be determined as
%$n-th$ round is 
    \begin{eqnarray}
f_n=\frac{1}{2}+\frac{5}{18}P\left(\lambda_1\right)P\left(\lambda_2\right).....P\left(\lambda_{n-1}\right)\lambda_n.
\end{eqnarray}
If we now assume that in each round, the lower bound of the fidelity is taken to be \(0.67\), i.e., just above the classical bound, the maximum attempting number in case of three receivers reduces to \emph{two} provided in the first round, all the receivers do not perform their unitary operations. 

%    \item 
\textbf{Theorem 1.} \emph{ No recycling is possible when the number of receivers exceeds three. }

\emph{Proof.} Analyzing   Eq. (\ref{fidelity for M receriver}), we realize that with the increase in the number of the receivers, the opportunity for recycling gradually decreases. And when the number of receiver  becomes four or more, i.e.,  $M\geq 4$,  leads to a condition on sharpness parameter which is unphysical in the second round.

Specifically, to derive the MAN for $M$ receivers, we set the fidelity in each round to be $f_c = \frac{2}{3}+\delta$, where $\delta$ is finite positive number close to zero for which we get quantum advantage in each round. Following Eq. (\ref{fidelity for M receriver}) in the first round, we have
\begin{eqnarray}
\nonumber f_1 &=& \frac{1}{2}+\frac{M+2}{6M}\lambda_1 = \frac{2}{3}+\delta,\\
\Rightarrow \lambda_1 &=& \frac{M(6\delta + 1)}{M+2}.
\end{eqnarray}
Similarly, in the second round, we obtain the fidelity as 
\begin{eqnarray}
\nonumber
f_2 &=& \frac{1}{2} + \frac{M+2}{6M}P(\lambda_1)\lambda_2 = \frac{2}{3}+\delta,\\ \nonumber
\Rightarrow \lambda_2 &=& \frac{12 M (\frac{1}{6}+\delta)}{2-6 M \delta + 2\sqrt{(1-3M\delta)(1+M(2+9\delta))}}.\\
\label{eq:f2}
\end{eqnarray}
To find whether it is possible to reuse the telecloning channel again, the condition on the sharpness parameter is
\begin{eqnarray}
\nonumber
\lambda_2 = \frac{12 M (\frac{1}{6}+\delta)}{2-6 M \delta + 2\sqrt{(1-3M\delta)(1+M(2+9\delta))}} \leq 1,\\
\end{eqnarray}
which gives the restriction on $\delta$ to be
\begin{eqnarray}
\delta \leq \frac{4-M}{18 M}.
\label{eq:proof_nogo}
\end{eqnarray}
From the above equation, it is evident that if the number of receivers, i.e., $M \geq 4$, we have $\delta \leq 0$. Therefore, we do not get any quantum advantage in the second round of telecloning when the number of receivers goes beyond $3$ and hence the proof.

$\blacksquare$
%\end{itemize}

\section{Consequence of disentangling operator on attempting telecloning }
\label{sec:nonoptimal}

Instead of using the optimal  state for telecloning, we start the protocol by taking non-optimal  state for telecloning as the shared multipartite resource state. Let us first introduce a disentanglement operator $\hat{D}$. The effect of this operator $\hat{D}_i$ on the  qubit $i$ in the computational basis is given by
\begin{eqnarray}
\nonumber
\hat{D}_i\ket{0}_i = \ket{0}_i,\,\, 
\hat{D}_i\ket{1}_i = \eta_i \ket{1}_i.
\end{eqnarray}
Notice that application of this operator on maximally entangled state, say on  $\ket{B_3} = 1/\sqrt{2}(\ket{01}+\ket{10})$ produces a non-maximally entangled state of the form, $\hat{D}_1 \ket{B_3} = 1/\sqrt{1+|\eta_1|^2}(\ket{01}+\eta_1\ket{10})$.

We now apply disentangling operator on each qubit of the two-receiver optimal telecloning state given in Eq. (\ref{optimalstate}) which modifies the state to \cite{rigolin07}
\begin{eqnarray}
\nonumber  \ket{\psi(\eta)}_{PAC_1 C_2} &=& B \bigg(\ket{0000}+\frac{\eta_P\eta_{C_1}}{2}\ket{1010}+\frac{\eta_A\eta_{C_1}}{2}\ket{0110}\\
\nonumber &+&\frac{\eta_P\eta_{C_2}}{2}\ket{1001}+\frac{\eta_A\eta_{C_2}}{2}\ket{0101}\\
&+&\eta_P\eta_A\eta_{C_1}\eta_{C_2}\ket{1111}\bigg),
\end{eqnarray}
where
\begin{eqnarray}
\nonumber B &=& \bigg(1+\frac{|\eta_P\eta_{C_1}|^2}{4}+\frac{|\eta_A\eta_{C_1}|^2}{4}+\frac{|\eta_P\eta_{C_2}|^2}{4}\\
&+&\frac{|\eta_A\eta_{C_2}|^2}{4}+|\eta_P\eta_A\eta_{C_1}\eta_{C_2}|^2 \bigg)^{-\frac{1}{2}}.
\end{eqnarray}
Here the set $\eta = \{\eta_P,\eta_A,\eta_{C_1},\eta_{C_2}\}$ contains all the disentangling parameters of the port, auxiliary qubit, $C_1$, and $C_2$. Note that $\eta_j$s can be taken to be real without loss of generality \cite{gordon06}, each of them varies from $0$ to $1$ and the unit values of all of them represent the optimal state. 

\begin{figure}
    \centering
    \includegraphics[width=0.9\linewidth]{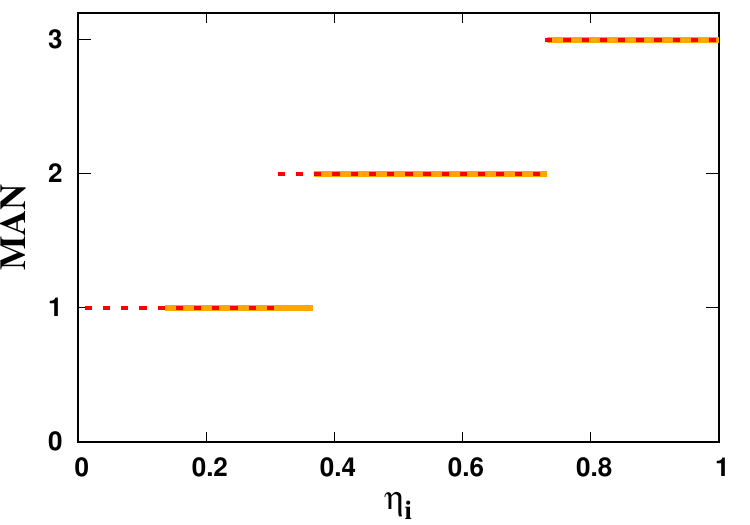}
    \caption{ \textbf{MAN vs. disentanglement parameter.} The dashed red line represents the situation with $\eta_i=\eta_{C_1(C_2)} = \eta_C$, while the orange solid line indicates the picture with $\eta_i=\eta_P$. In both the the cases, the maximal attempting number can reach three as obtained in the optimal telecloning case. Both the axes are dimensionless.  }
    \label{fig:disent}
\end{figure}

We now study the effect of non-optimal shared state on the maximal attempting number by considering different scenarios which emerge due to the different choices of  $\eta_j$s.
\begin{itemize}
    \item \textbf{Case $1$.} Let us take the situation, when $ \eta_P=\eta$ while $\eta_{C_1}=\eta_{C_2}=\eta_A=1 $, i.e., when only the port qubit is affected by the disentangling operator. By this action, one expects that  the entanglement of the shared state in the bipartition  \(P:AC_1C_2\) gets reduced, thereby decreasing the performance.  The fidelity of the telecloned state in the first round is given by
\begin{eqnarray}
\nonumber f_1
&=&\frac{1}{2}+\frac{1+4\eta+\eta^2}{9\left(1+\eta^2\right)}\lambda\\
&=&\left(\frac{1}{2}+\frac{\lambda}{9}\right)+\frac{4 \mathcal{C}\left(\eta\right)}{18}\lambda,
\end{eqnarray}
with $\mathcal{C}(\eta) = \frac{2 \eta}{1+\eta^2}$ is the concurrence \cite{wootters98} of $\ket{B_1}$ after applying the disentangling operator on the first party. %\textcolor{red}{B1 er concurrence mane? B1 to max ent state?
Notice that in this non-optimal channel, $f_1$ depends on the parameters of the initial state as 
\begin{eqnarray}
\nonumber
    f_1(\alpha,\beta) &=& \{2\alpha^2\beta^2(3+\eta^2(3-2\lambda)-2\lambda + 8 \eta \lambda) \\ &+& \nonumber \alpha^4 (5+\eta^2+5\lambda-\eta^2\lambda) \\\nonumber &+&\beta^4(1-\lambda+5\eta^2(1+\lambda))\}/\{6(1+\eta^2)\},\\
\end{eqnarray}
and hence the average fidelity is obtained as prescribed in Eq. (\ref{eq:1s2r}).

%In a similar fashion for optimal channel 
Following the same prescription as discussed in Sec. \ref{sec:result3}, we calculate the fidelity for the round $n$ provided all the Charus have not finished the protocol in previous $(n-1)$ rounds and it is given by
%which is found to be
\begin{eqnarray}
\nonumber    f_n &=&\left(\frac{1}{2}+\frac{P\left(\lambda_1\right)P\left(\lambda_2\right)....P\left(\lambda_{n-1}\right)}{9}\lambda_n\right)\\
    &+&\frac{4 \mathcal{C}\left(\eta\right)}{18}P\left(\lambda_1\right)P\left(\lambda_2\right)....P\left(\lambda_{n-1}\right)\lambda_n.
\end{eqnarray}
In Table. \ref{tab:etap},  we determine the range of $\eta$ when we choose $f_l = 0.67 >2/3$, by which the maximum  attempting number can be achieved. Interestingly, we find that even in presence of the disentangling operation which reduces the entanglement content of the shared state, there exists a range of \(\eta_P\) by which the maximal attempting number can still remain three as in the optimal shared state, reported in Sec. \ref{sec:result3}. 
\begin{table}[ht]
\centering
\begin{tabular}{|c|c|}
\hline
     Range of $\eta_P$ & MAN \\\hline
      $1-0.7327$& $3$\\\hline
      $0.7326- 0.3675$& $2$\\\hline
      $0.3674-0.1349$&$1$\\\hline
     % $0.13484-0$&$0$\\\hline
\end{tabular}
\caption{Range of $\eta_P$ with MAN, when $f_l=0.67$.}
\label{tab:etap}
\end{table}

    \item \textbf{Case $2$.} Let us now take $\eta_{C_1}=\eta_{C_2}=\eta_C$ and the rest of \(\eta_j\)s can be taken as unity, i.e.,  $\eta_P=\eta_A=1$. This scenario is in some sense complementary of Case 1 since the disentangling operation in this case acts on the receiver's end. It is interesting to find out which  disentangling operations (port or the receivers) have more adverse effects on MAN. 
    In the first round,  the optimal fidelity reads as
\begin{eqnarray}\label{eq1}
f_1=\frac{1}{2}+\frac{1}{6}\left(\frac{1+2\eta_C+2\eta_C^3+\eta_C^4}{1+\eta_C^2+\eta_C^4}\right)\lambda,
\end{eqnarray}
while in the  round $n$,  it becomes
\begin{eqnarray}
  \nonumber  f_n &=&\frac{1}{2}+\frac{1}{6}\left(\frac{1+2\eta_C+2\eta_C^3+\eta_C^4}{1+\eta_C^2+\eta_C^4}\right)\\
&\times&    P\left(\lambda_1\right)P\left(\lambda_2\right)\ldots P\left(\lambda_{n-1}\right)\lambda_n.
\end{eqnarray}
We compute  the ranges of $\eta_{C_1}= \eta_{C_2}=\eta_C$ (see Table. \ref{tab:etac}) for which the maximal attempting number remains constant to three by choosing \(f_l=0.67\). Comparing Tables \ref{tab:etap} and \ref{tab:etac}, we find that disentangling operation on port has much stronger consequence on the recycling of telecloning process compared to the disentangling operation on the receivers. 
In Fig. \ref{fig:disent},  we illustrate the maximal   attempting number for the entire range of $\eta_C$ and $\eta_P$ when $\lambda$  is fixed to $= 0.67$. In both the cases, the maximum rounds in which the protocol can be attempted with fidelity more than the classical one goes to three. 
%for the cases we discussed so far in the non-optimal scenario.

\begin{table}[ht]
\centering
\begin{tabular}{|c|c|}
\hline
     Range of $\eta_C$ & MAN \\\hline
      $1-0.7290$&$3$\\\hline
      $0.7289- 0.3115$&$2$\\\hline
      $0.3114-0.0101$&$1$\\\hline
 %     $0.01-0$&$0$\\\hline
\end{tabular}
\caption{By choosing \(f_l\) just above the classical fidelity, the range of  $\eta_{C_1}=\eta_{C_2} =\eta_C$ is listed against  MAN.}
\label{tab:etac}
\end{table}
\end{itemize}

%\item \textbf{Case $3$: If $\eta_{C_1}\neq\eta_{C_2} $ and $ \eta_P=\eta_A=1:$}}

%Fidelity in the nth step:
%\begin{eqnarray}
%\nonumber f_n = \frac{1}{2} &+&\frac{1}{3}\left[\frac{1+2\eta_{C_1}+2\eta_{C_1}\eta_{C_2}^2+\eta_{C_1}^2\eta_{C_2}^2}{2+\eta_{C_2}^2+2\eta_{C_1}^2+\eta_{C_1}^2\eta_{C_2}^2}\right]\\
%&\times & P\left(\lambda_1\right)P\left(\lambda_2\right)....P\left(\lambda_{n-1}\right)\lambda_n
%\end{eqnarray}

%\item {\textbf{Case $4$: If $\eta_{C_1}=\eta_{C_2}=\eta_C $, $ \eta_P\neq1$ and $ \eta_A=1:$}}\\
%Fidelity in the nth step:
%\begin{eqnarray}
%\nonumber f_n=\frac{1}{2}&+&\left[\frac{1+\eta_C\eta_P\left(2+2\eta_C^2+\eta_C^3\eta_p\right)}{6+3\eta_C^2\left(2\eta_C^2\eta_P^2+\eta_P^2+1\right)}\right]\\
%&\times& P\left(\lambda_1\right)P\left(\lambda_2\right)....P\left(\lambda_{n-1}\right)\lambda_n
%\end{eqnarray}

%\vspace{3cm}
In this non-optimal scenario, more asymmetry can also be introduced by applying different disentangling operations on different receivers, or in both the port and the receiver's ends and so on. For example, if $\eta_{C_1} \neq \eta_{C_2}$ and $\eta_P=\eta_A=1$,
the fidelity after step \(n\) can be written as
\begin{eqnarray}
\nonumber f_n = \frac{1}{2} &+&\frac{1}{3}\left[\frac{1+2\eta_{C_1}+2\eta_{C_1}\eta_{C_2}^2+\eta_{C_1}^2\eta_{C_2}^2}{2+\eta_{C_2}^2+2\eta_{C_1}^2+\eta_{C_1}^2\eta_{C_2}^2}\right]\\
&\times & P\left(\lambda_1\right)P\left(\lambda_2\right)....P\left(\lambda_{n-1}\right)\lambda_n,
\end{eqnarray}
 while the same reduces to  
\begin{eqnarray}
\nonumber f_n=\frac{1}{2}&+&\left[\frac{1+\eta_C\eta_P\left(2+2\eta_C^2+\eta_C^3\eta_p\right)}{6+3\eta_C^2\left(2\eta_C^2\eta_P^2+\eta_P^2+1\right)}\right]\\
&\times& P\left(\lambda_1\right)P\left(\lambda_2\right)....P\left(\lambda_{n-1}\right)\lambda_n
\end{eqnarray}
when $\eta_{C_1}=\eta_{C_2}=\eta_C $,  $ \eta_P\neq1$ and $\eta_A=1$. By choosing the disentangling parameters suitably, it is always possible to perform the telecloning protocol with quantum advantage for thrice provided all the receivers in previous rounds decline to complete the scheme. 

\section{Discussion}
\label{sec:discussion}

Quantum teleportation protocol which illustrates an infinite resource reduction using quantum mechanical systems is one of the main pillars in the field of communication.
In particular, to send an arbitrary qubit from a sender to a receiver,
quantum protocol  requires only two bits of classical communication  provided an entangled resource state is shared between them, instead of an infinite amount of classical communication. After the discovery of the theoretical protocol, it has been experimentally verified in several physical systems like photons, continuous variable systems, ion traps etc and has also been extended in several directions.   One of the interesting avenues in the field of quantum communication is to build a quantum network involving multiple senders and multiple receivers. In this direction, it was shown that instead of sharing multiple maximally entangled states,  genuine multipartite entangled states can have some beneficial role in multipartite quantum communication protocols. For example,  Greenberger-Horne-Zeilinger (GHZ) state \cite{greenberger1989} or W state \cite{DVC00} are found to be useful for both multipartite version of quantum teleportation and dense coding \cite{hhhh, bruss04, agarwal06}. 
%can be used for teleportation but the prototype W-states can't. For two sender and single receiver picture GHZ state can give advantage over multiple shared Bell states. But its not advantageous to use it in case of single sender and two receiver. Taking the telecloning state as resource $N$ senders can send an unknown state to $M (\geq N) $ spatially separated receivers with optimal fidelity, upper bounded by consistency of quantum mechanics. 

A prominent example of teleportation in a multipartite domain include the telecloning protocol where a single sender wants to send an arbitrary qubit to multiple receivers, and hence this protocol is restricted by the bounds obtained via the approximate cloning machine. Specifically, optimal universal cloning machine incorporates  a multiparty  entangled state such that the fidelity of the output cloned state with the input state is independent of the initial choice of the state \cite{cloning98} and it was shown that  by choosing that multiparty entangled state as the quantum channel, one can design a teleportation protocol having a single    sender and multiple receivers. Interestingly, one can find that  instead of sharing multipartite entangled states like the GHZ or the W states, the protocol is successful when an optimal state obtained from the cloning machine is shared as a quantum channel. Telecloning protocol originally uses projective sharp Bell measurement, same as all previous teleportation protocols. After the measurement, the correlation between the sender and reciever's side is destroyed leading to nonreusability  of the quantum channel for further communication tasks.

 Role of unsharp measuremnt in  sequential scenarios are already explored in different directions, although they are mostly restricted to the identification  of entangled states or detection of nonlocality etc. In this work, we go beyond the detection of entanglement and illustrate the usefulness of unsharp measurement in the multipartite quantum communication protocol, i.e, in the telecloning protocol. 
 In particular, unlike projective Bell measurement, the measurement at the sender's side in telecloning is made unsharp so that the quantum correlations between the sender and the receivers do not get destroyed after the measurement and hence the shared state remains useful for some quantum information tasks even after a few rounds of the protocol. Our aim is to design a protocol involving unsharp measurement and multipartite quantum channel so that quantum communication protocol can be performed in any step of the scheme. Further, we are interested to find the number of reusability of a given quantum channel for telecloning.
 
 %In our work, we investigated the telecloning scheme in the context of unsharp measurement sequential scenario where the initial quantum channel can be reused in different instances leading to a quantum telecloning network. 

Suppose after the first round, receivers are unable to perform the unitary operations for some reasons, and   the resulting entangled state due to the unsharp measurement can then be reused for another round of the protocol for some suitable range of unsharp parameters. We proved that  when the shared state for telecloning is the optimal one, the maximum round in which the protocol can be performed with quantum advantage is three. We found that the maximum attempt number (MAN), i.e., the maximum number of rounds in which the mentioned quantum channel can be reused with quantum advantage, decreases with the increase in critical fidelity  which we demand in each round. Even with the shared nonoptimal multipartite states obtained after applying disentangling operations at the sender's or the receivers' sides or both of the optimal state, we reported the MAN to be three for some range of disentangling parameters. We showed that the fidelity obtained in each round is connected with the entanglement content of the reduced shared channel between the sender and the concerned receiver as well as with the monogamy score of the channel between the sender and the receivers. Specifically, maximum achievable fidelity in the second round goes below the classical fidelity of $\frac{2}{3}$, when the entanglement  in the bipartition vanishes. We also demonstrated that the reattempting scenario is meaningful in the telecloning scheme only when it involves a single sender and the maximum three receivers, thereby arriving to a no-go theorem on  the reusability of the quantum channel in telecloning scheme.

Our results demonstrate  the power of unsharp measurements in quantum communication protocol which possibly indicates the importance of  designing quantum information processing tasks based on unsharp measurements.

\section*{acknowledgements}

We acknowledge the support from Interdisciplinary Cyber Physical Systems (ICPS) program of the Department of Science and Technology (DST), India, Grant No.: DST/ICPS/QuST/Theme- 1/2019/23. We  acknowledge the use of \href{https://github.com/titaschanda/QIClib}{QIClib} -- a modern C++ library for general purpose quantum information processing and quantum computing (\url{https://titaschanda.github.io/QIClib}) and cluster computing facility at Harish-Chandra Research Institute. 

\appendix
%\label{appendixa}
\section{Optimal state evolution}

The density matrix corresponds to the optimal state in Eq. (\ref{optimalstate}) is given by
\begin{eqnarray}
\rho_{PAC}=\ket{\psi}_{PAC_1 C_2}\bra{\psi}.
\end{eqnarray}
After performing unsharp  POVM measurements \(n\) times, the recycled state at round $n$ can be written as
\begin{eqnarray}
\nonumber\rho_{PAC_1C_2}^n=p_n\ket{\psi}_{PAC}\bra{\psi}+\left(\frac{1-p_n}{6}\right)\Bigg[\frac{{\mathbb{I}}_4}{2}\otimes\ket{B_3}\bra{B_3}\\
\nonumber +\mathbb{I}_2\otimes\Big(\ket{000}\bra{000}+\ket{111}\bra{111}\Big)\\
\nonumber +\frac{\mathbb{I}_2}{2}\otimes\Big(\ket{0}\bra{1}\otimes\mathbb{I}_2\otimes\ket{0}\bra{1}+\ket{1}\bra{0}\otimes\mathbb{I}_2\otimes\ket{1}\bra{0}\\
 \nonumber+ \ket{0}\bra{1}\otimes\ket{0}\bra{1}\otimes\mathbb{I}_2+\ket{1}\bra{0}\otimes\ket{1}\bra{0}\otimes\mathbb{I}_2\Big)\Bigg]\\
\end{eqnarray}

where $p= P\left(\lambda_1\right)P\left(\lambda_2\right).....P\left(\lambda_{n}\right)$ and $n=\{1,2\}$. 

\section{Logarithmic Negativity for 2-nd recycled channel}
\label{sec:logneg}

%\begin{itemize}
Let us first give the definition of logarithmic negativity to quantify entanglement in a bipartite state, \(\rho_{AB}\). Based on it, we also compute the monogamy score of entanglement whose definition will also be given below.     
For any operator,  the trace norm can be calculated as $\left \| A  \right \| = \sqrt{tr(A^{\dagger}A)}$ which is basically the sum of the singular values. We can define negativity, a non-convex entanglement monotone \cite{vidal2002, plenio2005} as
\begin{eqnarray}
\mathcal{N}(\rho_{A:B})= \frac{\left \| \rho^{\Gamma_{A}} \right \| -1 }{2}
\end{eqnarray}
where $\rho^{\Gamma_{A}}$ is the partial transpose of $\rho_{A:B}$ with respect to party \(A\) \cite{peres, Horodecki96}. Using this, logarithmic negativity  is defined as
\begin{eqnarray}
LN(\rho_{A:B}) = \log{\left \| \rho^{\Gamma_{A}} \right \|}= \log(2\mathcal{N}+1),
\label{eq:LN}
\end{eqnarray}
which reduces to the modulus of a negative eigenvalue in a two-qubit case.  

To relate  entanglement in the multipartite  channel with its reusability and the maximum achievable fidelity in each round, we  calculate $LN(\rho_{A:B})$ between  Alice (P) and one of the receivers $C_{1}(C_{2})$. For the first recycled channel, we compute logarithmic negativity which is given in Eq. (\ref{eq:LN1st}) while the same for the second recycled state can be found to be 
%for different sender-reciever bipartition has been expressed in the main text in terms of predecided fidelity of first round. Similarly,for second recycled quantum channel , logarithmic negativity for different bipartition   can be written in terms of fidelity of first and second round($f_{1},f_{2}$) as
\begin{widetext}
\begin{eqnarray}
   \nonumber LN_{P:C_1(C_2)}&=&\log_2\bigg[\frac{1}{12}\bigg|3.5+3f_1-X_1+\bigg(3f_1-2.5-X_1\bigg)\bigg(X_2-X_3\bigg)\\
    &-&4\sqrt{(3f_1-2.5)(3f_1-0.5+X_1)\Big(X_3-1\Big)\Big(1+X_2+X_3\Big)}\bigg|+1\bigg].
    \label{eq:LN2}
      \end{eqnarray}
      \end{widetext}
The behavior of the above expression is plotted in Fig. \ref{fig:2nd recycle} (b).

\section{Monogamy score}
\label{sec:monogamy}

In contrast
to classical correlations, quantum correlations cannot be shared arbitrarily among parties in a multipartite state.  Specifically, in a tripartite state, \(\rho_{ABC}\), if \(A\) and \(B\) are highly entangled, monogamy of entanglement says that the entanglement content between \(A\) and \(C\) cannot be large \cite{coffman2000, ckw2,  monorev}.  
%, , satisfy a monogamy relation. If this kind of monogamy
%relation is  satisfied by certain measures of entanglement, so that if two parties are highly entangled, then
%they cannot have a large amount of entanglement shared
%with a third one . 
More precisely, the
monogamy inequality for a bipartite quantum correlation
measure, $\mathcal{Q}$, for a \(N\)-party state,  $\rho_{A_1 A_2 \ldots A_N}$ can be written as
\begin{equation}
   \sum_i \mathcal{Q}(\rho_{A_1: A_i})  \leq \mathcal{Q}(\rho_{A:A_2 \ldots A_N}). 
    \label{eq:monogamy}
\end{equation}
Based on this inequality, one can define a shareability measure of entanglement, known as  monogamy score of entanglement  as
\begin{equation}
    \delta_{\mathcal{Q}} = \mathcal{Q}(\rho_{A:A_2 \ldots A_N}) -  \sum_i \mathcal{Q}(\rho_{A_1: A_i}).  
    \label{eq:monogamysc}
\end{equation}
%Such that any tripartite state having positive monogamy score satisfies monogamy relation and negative monogamy score violates monogamy relation.
In the paper, we choose LN as a measure of entanglement which we denote it as \(\delta_{LN}\). 

In the telecloning protocol involving a single sender, two receivers and an auxiliary qubit, the summation in \(\delta_{LN}\) contains three terms. Among them, \(LN(\rho_{P:A}) =0\) while \(LN_{PC_1 (C_2)}\) are calculated in Eqs. (\ref{eq:LN1st}) and (\ref{eq:LN2}). And the first term after the second and third round read respectively as
     %\end{widetext}
    \begin{eqnarray}
   \nonumber && LN_{P:AC_1C_2}=\\
  \nonumber  &&\log_2(\frac{1}{4}|(-0.5+3f_1-\sqrt{2.5-3f_1}\sqrt{9f_1-3.5}-2\sqrt{2}\\
    \nonumber &\times&\sqrt{(2.5-3f_1)(3f_1-0.5+\sqrt{2.5-3f_1}\sqrt{9f_1-3.5})})|\\
    &+&1),
    \end{eqnarray}
    and      
\begin{widetext}
    \begin{eqnarray}
  \nonumber LN_{P:AC_1C_2}&=&\log_2\bigg[\frac{1}{8}\bigg|1.5+3f_1-X_1+\bigg(3f_1-2.5-X_1\bigg)\bigg(X_2-X_3\bigg)\\
    &-&4\sqrt{(3f_1-2.5)(3f_1-0.5+X_1)\Big(X_3-1\Big)\Big(1+X_2+X_3\Big)}\bigg|+1\bigg],
    \end{eqnarray}
where  
\begin{eqnarray}
X_1&=&\sqrt{2.5-3f_1}\sqrt{9f_1-3.5},\\
X_2&=&\sqrt{1-\frac{3f_2-1.5}{P\left(f_1\right)}}\sqrt{1+\frac{9f_2-4.5}{P\left(f_1\right)}},\\
X_3&=&\frac{3f_2-1.5}{P\left(f_1\right)},\\
P\left(f_1\right)&=&\frac{1}{4}\Big(5-6f_1+\sqrt{(5-6f_1)(18f_1-7)}\Big).
\end{eqnarray}

     \end{widetext}

%\vspace{1cm}

\bibliography{ref}

\end{document}